\documentclass[conference]{IEEEtran}
\IEEEoverridecommandlockouts
\ifCLASSINFOpdf
\else
\usepackage[dvips]{graphicx}
\fi
\usepackage{url}

\usepackage{cite}
\usepackage{graphicx,amsmath,amssymb,tikz,psfrag}
\usepackage{color}
\usepackage{fancyhdr} 
\usepackage{tikz}
\usetikzlibrary{shapes,arrows,chains}
\usepackage{xcolor}
\usepackage{algorithm}
\usepackage{algpseudocode}
\usepackage{enumerate}
 
\begin{document}

	\title{\huge The Generalized Degrees-of-Freedom of the Asymmetric Interference Channel with Delayed CSIT}
	
	\author{\IEEEauthorblockN{Tong Zhang\textsuperscript{$1$}, Yufan Zhuang\textsuperscript{$1$}, and Yinfei Xu\textsuperscript{$2$},} 
		%Junge Wang, Na Li, Rui Wang, and Xiaofeng Tao %and Rui Wang,%and Tobias J. Oechtering
		
	\IEEEauthorblockA{\textsuperscript{$1$}Department of Electrical and Electronic Engineering, Southern University of Science and Technology, Shenzhen, China}
	%\IEEEauthorblockA{\textsuperscript{$2$}Center for Pervasive Communications and Computing (CPCC), University of California, Irvine, USA}
	\IEEEauthorblockA{\textsuperscript{$2$}School of Information Science and Engineering, Southeast University, Nanjing, China}
	
	Email: \{zhangt7, zhuangyf2019\}@sustech.edu.cn, yinfeixu@seu.edu.cn

			 	 %\thanks{This work was supported in part by  the China Postdoctoral Science Foundation under Grant 2021M691453, in part by the National Natural Science Foundation of China under Grant 61901105 and 62171119, in part by the Natural Science Foundation of Jiangsu Province under Grant BK20190343, in part by the Zhi Shan Young Scholar Program of Southeast University.}
 
	}\maketitle

	\begin{abstract}
 In this paper, we investigate the generalized degrees-of-freedom (GDoF) of the asymmetric interference channel with delayed channel state information at the transmitter (CSIT), where each transmitter has two antennas, each receiver has one antenna, and the strength for each interfering link can vary. The optimal sum-GDoF is characterized by matched converse and achievability proof. Through our results, we also reveal that in our antenna setting, the  symmetric GDoF lower bound in [Mohanty et. al, TIT 2019]  can be elevated, and the symmetric GDoF upper bound in [Mohanty et. al, TIT 2019] is tight in fact. 
	\end{abstract}
	
	
	\section{Introduction}
	
	The degrees-of-freedom (DoF) characterization with delayed channel state information at the transmitter (CSIT) has attracted a plenty of research interests in the past decade. For example, the DoF region of multiple-input multiple-output (MIMO) interference channel with delayed CSIT was derived in \cite{24}. The study of DoF of MIMO broadcast channel with delayed CSIT can be found in \cite{25,26,27,28}, whose exact value was still not completely obtained. The linear DoF region of MIMO X channel with delayed CSIT was characterized in \cite{29}.
	
	One limitation of DoF is that the strength of each link is assumed to be equal, whereas the link strength can vary tremendously in wireless. Nevertheless, the generalized degrees-of-freedom (GDoF) overcomes this drawback by considering the different strength of each link, which was first proposed in \cite{40}.  The GDoF characterization with delayed CSIT can be found in \cite{31,30,20,21}.  In \cite{31}, the GDoF was studied in the two-user multiple-input single-output (MISO) broadcast channel under alternating delayed, perfect, and no CSIT, where sum-GDoF upper and lower bounds were shown to be partially coincided. In \cite{30}, the secure GDoF was investigated in the two-user MISO broadcast channel with an external eaversdropper and alternating delayed, perfect, and no CSIT. The GDoF region of the MIMO Z channel with delayed CSIT was characterized in \cite{20}. For MIMO interference channel with delayed CSIT and symmetric interfering link strengths, symmetric GDoF upper and lower bounds were derived in \cite{21}, where the symmetric GDoF was characterized partially, and the  upper and lower bounds do not change with antenna ratio, i.e.,  number of antenna at each transmitter over number of  antennas at each receiver, if the antenna ratio is equal to or larger than two\footnote{A representative antenna setting for this antenna configuration case is that the transmitter has 2 antennas and the receiver has 1 antenna.}. 
	
	In this paper, we  study the GDoF of the asymmetric interference channel with delayed CSIT, where the transmitter has two antennas, the receiver has one antenna, and the strength for each interfering link can vary. We characterize the sum-GDoF by presenting matched converse and achievability proof. Then, via this result, we reveal that in our antenna setting, the  symmetric GDoF lower bound in \cite{21}  can be elevated, and the symmetric GDoF upper bound in \cite{21} is tight in fact. The idea of elevating the GDoF lower bound comes from \cite{30}, where the authors in \cite{30} fully exploit the receiver signal space by sending new fresh symbols in each time slot. Our transmission scheme generalizes the  scheme  in \cite[Appendix C]{30}, which was designed for achieving the corner points of GDoF region of MISO broadcast channel with delayed CSIT.

	%\textit{Notations}:
	
	\begin{figure}
		\centering
		\includegraphics[width=2.4in]{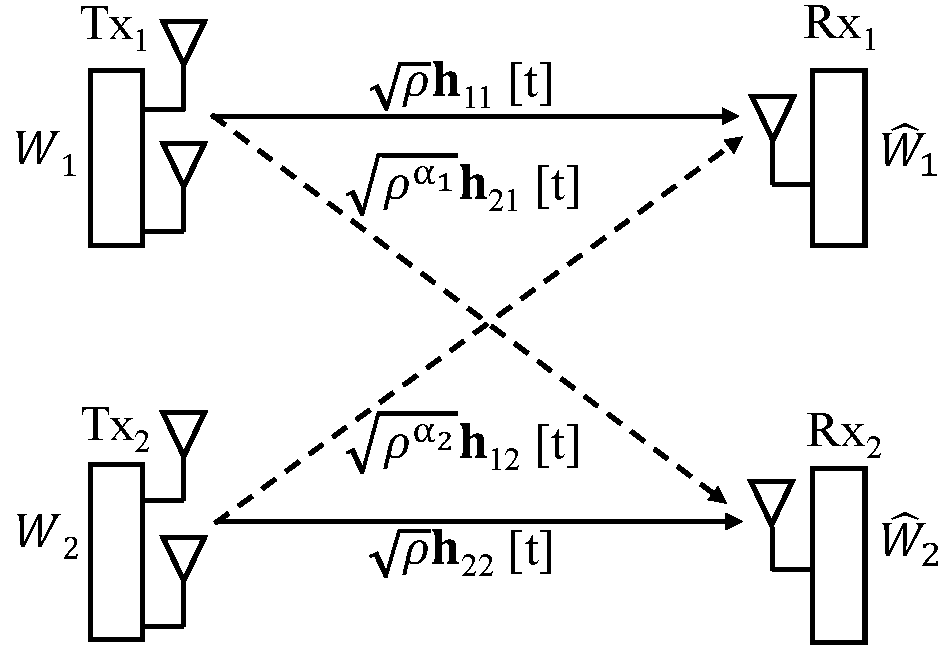}
		\caption{The considered scenario of asymmetric interference channel, where each transmitter has 2 antennas and each receiver has 1 antenna.}
	\end{figure}
	
	\section{System Model}

	The considered asymmetric interference channel has two transmitters, denoted by $\text{Tx}_1$ and $\text{Tx}_2$, each with two antennas, and two receivers, denoted by $\text{Rx}_1$ and $\text{Rx}_2$, each with one antenna, as depicted in Fig. 1. The transmitter $\text{Tx}_i,\,i=1,2,$ sends private message $W_i$ to the receiver $\text{Rx}_i$. The received signals at two receivers and time slot $t$ can be written as
	\begin{subequations}
		\begin{eqnarray}
	 	y_1[t] = \sqrt{\rho}\textbf{h}_{11}[t]\textbf{x}_1[t] + \sqrt{\rho^{\alpha_2}}\textbf{h}_{12}[t]\textbf{x}_2[t] + n_1[t], \\
		 	y_2[t] =  \sqrt{\rho^{\alpha_1}}\textbf{h}_{21}[t]\textbf{x}_1[t] + \sqrt{\rho}\textbf{h}_{22}[t]\textbf{x}_2[t] + n_2[t],
		\end{eqnarray}
	\end{subequations}
	where $\textbf{x}_i[t] \in \mathbb{C}^{2 \times 1}$ denotes the transmitted signal from transmitter $\text{Tx}_i$ at time slot $t$, $y_j[t] \in \mathbb{C}^{1},\,j=1,2,$ denotes the received signal at receiver $\text{Rx}_j$ and time slot $t$, $\textbf{h}_{ji} \in \mathbb{C}^{1 \times 2}$ denotes the channel matrix from transmitter $\text{Tx}_i$ to receiver $\text{Rx}_j$, $n_j \sim {\cal{CN}}(0,\sigma^2)$ denotes the additive White Gaussian noise (AWGN) at receiver $\text{Rx}_j$, the channel gains of desired links and interfering links are denoted by $\sqrt{\rho}$ and $\sqrt{\rho^{\alpha_i}}$ with $0 < \rho$ and $0 \le \alpha_i$, respectively. Without loss of generality, we assume that $\alpha_2 \le \alpha_1$. The transmit signals follow an average power constraint, given by $\frac{1}{n} \sum_{t=1}^n \text{tr}\left( \mathbb{E}\{\textbf{x}_i[t]\textbf{x}_i[t]^H\}\right) \le 1$. All entries of channel matrices are independent and identical distributed (i.i.d.) across space and time slot. The signal-to-noise ratio (SNR) at each receiver is $\rho$, and interference-to-noise ratio (INR) at receivers $\text{Rx}_1$ and $\text{Rx}_2$ are $\rho^{\alpha_2}$ and $\rho^{\alpha_1}$, respectively. We define the following assembly of channel matrices: $\mathcal{H}[t] \triangleq \{\textbf{h}_{ji}[t]\}_{i,j=1,2}$, and $\mathcal{H}^\tau \triangleq \{\mathcal{H}[t]\}_{t=1}^\tau$.
	
	Due to feedback delay, the transmitter has delayed channel state information. Specifically, at time slot $t$, the transmitters know $\mathcal{H}^\tau$, namely all the channel matrices up to time slot $t-1$. Two receivers have instantaneous knowledge of channel matrices. The encoding function at transmitter $\text{Tx}_i$ and time slot $t$ is denoted by $e_{i,t}(W_i,\mathcal{H}^{t-1})$. The decoding function at receiver $\text{Rx}_j$ after $n$ time slots is denoted by $c_{j,n}(W_i,\mathcal{H}^n)$.
	
	The rate tuple is written as $(R_1(\rho,\alpha_1,\alpha_2),R_2(\rho,\alpha_1,\alpha_2))$, where rate $R_i = \frac{\log|\mathcal{W}_i|}{n}$ is the cardinality of message set $\mathcal{W}_i$. The rate is achievable, if there are a sequence of codebook pairs $\{\mathcal{C}_{1,t},\mathcal{C}_{1,t}\}_{t=1}^n$ and decoding functions $\{c_{1,n},c_{2,n}\}$ such that the error probability $P_{e}^{(n)}$ goes to zero when $n$ goes to infinity. The sum-capacity is defined as  the supremum of sum of achievable rates, i.e., $C_\text{sum} = \sup \sum_{i=1}^2 R_i(\rho,\alpha_1,\alpha_2)$.
	%Then, the GDoF region $\mathcal{D}(\alpha_1,\alpha_2)$ is defined as the pre-log factor of the capacity region as $\rho$ goes to infinity, i.e., 
	%\begin{eqnarray}
	%	\mathcal{D}(\alpha_1,\alpha_2) = \left\{ (d_1(\alpha_1,\alpha_2),d_2(\alpha_1,\alpha_2)\left|\right.0 \le d_i(\alpha_1,\alpha_2),  \right. \nonumber\\ 
	%	\qquad \text{and}\,\, (R_1(\rho,\alpha_1,\alpha_2),R_2(\rho,\alpha_1,\alpha_2)) \in \mathcal{C}(\rho,\alpha_1,\alpha_2),\nonumber\\ 
	%	\text{such that} \,\,d_i = \lim_{\rho \rightarrow {\cal{1}}} \frac{R_i(\rho,\alpha_1,\alpha_2)}{\log \rho},\,\,i=1,2.\}
	%\end{eqnarray}
Then, the sum-GDoF is defined as the pre-log factor of sum-capacity, i.e.,
$
	\sum_{i=1}^2 d_i = \lim_{\rho \rightarrow {\cal{1}}} \frac{C_\text{sum}}{\log \rho}.
$

	\section{Main Results and Discussion}

\textbf{Theorem 1}: For the considered asymmetric interference channel with delayed CSIT, defined in Section-II, the sum-GDoF is given as follows:
\begin{eqnarray}
	 \sum_{i=1}^2 d_i  = \begin{cases}
		2 - \dfrac{\alpha_1 + \alpha_2}{3} , & \alpha_1,\alpha_2 \le 1, \\
			 \min \left\{\dfrac{4 + \alpha_1 - \alpha_2}{3},2\right\}, & \begin{aligned}
			& 1 < \alpha_1\,\&\,\alpha_2 \le 1  \\
			& \&\, 2 \le \alpha_1 + 2\alpha_2 
			 \end{aligned},  \\
		\min\left\{\dfrac{2+\alpha_1 + \alpha_2}{3},2\right\}, & 1 < \alpha_1,\alpha_2.		
		\end{cases} \label{BE10}
\end{eqnarray}
 	
\begin{IEEEproof}
	Please refer to Section-IV for the converse proof and Section-V for the achievability proof.
	\end{IEEEproof}

%\textbf{Theorem 2}: For the considered asymmetric interference channel with delayed CSIT, defined in Section-II, the sum-GDoF lower bound is given as follows:

\textbf{Remark 1}: This sum-GDoF in \eqref{BE10} degenerates to optimal sum-DoF in \cite{24}, by setting $\alpha_1 =\alpha_2 = 1$, whose value is the celebrated $4/3$. Furthermore, in our antenna setting, it can be verified that the  symmetric GDoF upper bound in \cite{21} is tight. 

%\textbf{Theorem 2}: For the considered asymmetric interference channel with delayed CSIT,  defined in Section-II, the lower bound of sum-GDoF is given as follows:
%\begin{eqnarray}
%	&& \sum_{i=1}^2 d_i  \ge \nonumber \\
%	&& \begin{cases}
%	2 - \dfrac{\alpha_1 + \alpha_2}{3}, & \alpha_1,\alpha_2 \le 1 \,\&\, \alpha_1 \le 2 \alpha_2, \\
%	\dfrac{4 + \alpha_1 - \alpha_2}{3}, & 1 < \alpha_1\,\&\,\alpha_2 \le 1,  \\
%		\min\left\{\dfrac{2+\alpha_1 + \alpha_2}{3},2\right\}, & 1 < \alpha_1,\alpha_2, 		
%	\end{cases}  
%\end{eqnarray}

%\begin{IEEEproof}
%	Please refer to Section-V.
%\end{IEEEproof}
%\newpage
 
%\textbf{Remark 1}:
 
	\section{Converse Proof of Theorem 1}
	
	The key steps of this proof follows the that in \cite{21}. To begin with, we define the following virtual received signals, which are obtained from removing the impact of $\textbf{x}_1[t]$ in the received signals at each receiver:
	\begin{subequations}
		\begin{eqnarray}
		 \overline{y}_1[t] = \sqrt{\rho^{\alpha_2}}\textbf{h}_{12}[t]\textbf{x}_2[t] + n_1[t], \\
		  \overline{y}_2[t] = \sqrt{\rho}\textbf{h}_{22}[t]\textbf{x}_2[t]  + n_2[t]. 
		\end{eqnarray}
	\end{subequations}
	Before the next step, we define the following assembly of channel matrices: $\overline{\mathcal{Y}}_i^\tau \triangleq \{\overline{y}_i[t]\}_{t=1}^\tau$, $\mathcal{Y}_i^\tau \triangleq \{y_i[t]\}_{t=1}^\tau$, and $\mathcal{X}_i^\tau \triangleq \{\textbf{x}_i[t]\}_{t=1}^\tau$, $i=1,2$. Since the error probability $P_e^{(n)}$ goes to zero as $n$ goes to infinity, we denote $n\epsilon_n \triangleq 1 + n R_i P_{e}^{(n)}$ so that $\lim_{n \rightarrow {\cal{1}}} \epsilon_n = 0$. According to \cite{21}, the rate of receiver 1 can be bounded as
	\begin{eqnarray}
			  n(R_1 - \epsilon_n)  \le \sum_{t=1}^n h(y_1[t]|\mathcal{H}[t]) - \sum_{t=1}^n h(\overline{y}_1[t]|\mathcal{U}[t],\mathcal{H}[t]), \label{BE2}
	\end{eqnarray}
where $\mathcal{U}[t] \triangleq \{\overline{\mathcal{Y}}_1^{t-1},\overline{\mathcal{Y}}_2^{t-1},\mathcal{H}^{t-1}\}$. Next, according to \cite{21}, the rate of receiver 2 can be bounded as 
	\begin{eqnarray}
 n(R_2 - \epsilon_n) \le \sum_{t=1}^n h(\overline{y}_1[t],\overline{y}_2[t]|\mathcal{U}[t],\mathcal{H}[t]).  \label{BE3}
	\end{eqnarray}	
	
	Henceforth, we define 
 $\textbf{S}[t] \triangleq  [\sqrt{\rho^{\alpha_2}} \textbf{h}_{12}[t], 
				\sqrt{\rho} \textbf{h}_{22}[t]]^T$, 
	 	$\textbf{K}[t] \triangleq \mathbb{E}\{\textbf{x}_2[t]\textbf{x}_2[t]^H|\mathcal{U}[t]\}$,  
		 $\textbf{L}[t] \triangleq \mathbb{E}\{\textbf{x}_1[t]\textbf{x}_1[t]^H|\mathcal{H}^{t-1}\}$,
	  $\mathcal{V}[t] \triangleq \{\mathcal{U}[t],\mathcal{H}[t]\}$, 
where the transmit covariance matrix $\textbf{K}[t]$ is independent of $\textbf{h}_{12}[t],$  $\textbf{h}_{22}[t]$, and $\textbf{S}[t]$. Applying the extremal inequality in \cite{32} for the physically degraded channel $\mathcal{X}_2^\tau \rightarrow (\overline{\mathcal{Y}}_1^n,\overline{\mathcal{Y}}_2^n) \rightarrow \overline{\mathcal{Y}}_1^n$, we have the following inequality:
\begin{eqnarray}
	&& \frac{h(\overline{y}_1[t],\overline{y}_2[t]|\mathcal{V}[t])}{2} - h(\overline{y}_1[t]|\mathcal{V}[t]) \nonumber \\
	&& \le \max_{\begin{matrix}
			\textbf{K}[t] \succeq 0 \\
			\text{tr}\{\textbf{K}[t]\} \le 1
		\end{matrix}} \mathbb{E} \left\{ \log \left| \textbf{I}_2 + \textbf{S}[t]\textbf{K}[t]\textbf{S}[t]^H\right|/2 \right. \nonumber \\
	&& \left. \quad - 	\log \left| 1 + \rho^{\alpha_2} \textbf{h}_{12}[t]\textbf{K}[t]\textbf{h}_{12}[t]^H \right|
	\right\}. \label{BE1} 
\end{eqnarray}
To proceed, we can approximate the first term of \eqref{BE1} as  
\begin{eqnarray}
	&& \!\!\!\!\!\!\!\!\!\!\! \log \left| \textbf{I}_{2} + \textbf{S}[t]\textbf{K}[t]\textbf{S}[t]^H\right| \nonumber \\
	&& \!\!\!\!\!\!\!\!\!\!\! \overset{(a)}{=} \log \left| \textbf{I}_2 + \begin{bmatrix}
		\sqrt{\rho^{\alpha_2}} \widetilde{\textbf{h}}_{12}[t] \\
		\sqrt{\rho} \widetilde{\textbf{h}}_{22}[t]
	\end{bmatrix} 
\begin{bmatrix}
	\sqrt{\rho^{\alpha_2}} \widetilde{\textbf{h}}_{12}[t] \\
	\sqrt{\rho} \widetilde{\textbf{h}}_{22}[t]
\end{bmatrix}^H
\right| \nonumber \\
&& \!\!\!\!\!\!\!\!\!\!\! \overset{(b)}{=} \log  \left| \textbf{I}_{2-k_t} + \rho^{\alpha_2}\widetilde{\textbf{h}}_{12}[t]^H\widetilde{\textbf{h}}_{12}[t] +  \rho\widetilde{\textbf{h}}_{22}[t]^H\widetilde{\textbf{h}}_{22}[t]\right| \nonumber \\
&& \!\!\!\!\!\!\!\!\!\!\! \overset{(c)}{=} f(2-k_t, (\alpha_2,1), (1,1)) \log \rho + \mathcal{O}(1) \nonumber \\
&& \!\!\!\!\!\!\!\!\!\!\! = \begin{cases}
(\min\{2-k_t,1\} + \\
\quad \min\{[1 - k_t]^+, 1\}\alpha_2)	\log \rho + \mathcal{O}(1), & \alpha_2 \le 1,\\
(\min\{2-k_t, 1\}\alpha_2 + \\
\quad \min\{[1 - k_t ]^+,1\})	\log \rho + \mathcal{O}(1), & 1 < \alpha_2,
\end{cases} \label{BE6}
\end{eqnarray}	
where (a) is from SVD of $\textbf{K}[t]$, i.e., $\textbf{K}[t] = \textbf{U}[t]{\bf{\Sigma}}[t]\textbf{U}[t]^H$ with unitary matrix $\textbf{U}[t] \in \mathbb{C}^{2 \times (2-k_t)}$ and diagonal matrix ${\bf{\Sigma}}[t] \in \mathbb{C}^{(2-k_t)\times(2-k_t)}$, and $\widetilde{\textbf{h}}_{j2}[t] \triangleq \textbf{h}_{j2}[t]\textbf{U}[t]{\bf{\Sigma}}[t]^{1/2}$, where $k_t \in \{0,1,2\}$ denotes the number of zero singular values; (b) is from $|\textbf{I} + \textbf{AB}| = |\textbf{I} + \textbf{BA}|$; (c) is from \cite[Lemma 1]{21}.
The second term of \eqref{BE1} can be approximated as 
\begin{eqnarray}
	&& \log \left| 1 + \rho^{\alpha_2} \textbf{h}_{12}[t]\textbf{K}[t]\textbf{h}_{12}[t]^H \right| \nonumber \\
	&& \overset{(a)}{=}  \min\{2-k_t,1\} \alpha_2 \log \rho + \mathcal{O}(1), \label{BE7}
\end{eqnarray}
where (a) is from SVD of $\textbf{K}[t]$ and \cite[Lemma 1]{21}. Next, we can approximate the first term of \eqref{BE2} as 
\begin{eqnarray}
	&& \!\!\!\!\!\!\!\!\!\!\!  h(y_1[t]|\mathcal{H}[t]) \nonumber \\
	&& \!\!\!\!\!\!\!\!\!\!\! \overset{(a)}{\le} \log \left| 1 + \rho\textbf{h}_{11}[t]\textbf{L}[t]\textbf{h}_{11}[t]^H+ \rho^{\alpha_2}\textbf{h}_{12}[t]\textbf{K}[t]\textbf{h}_{12}[t]^H\right| \nonumber \\
	&& \!\!\!\!\!\!\!\!\!\!\! \overset{(b)}{\le} \log \left|1 + \rho \textbf{h}_{11}[t]\textbf{h}_{11}[t]^H +  \rho^{\alpha_2} \widetilde{\textbf{h}}_{12}[t]\widetilde{\textbf{h}}_{12}[t]^H \right| \nonumber \\
	&& \!\!\!\!\!\!\!\!\!\!\! \overset{(c)}{=}  f(1,(1,2),(\alpha_2,2-k_t)) \log \rho + \mathcal{O}(1), \nonumber \\
	&& \!\!\!\!\!\!\!\!\!\!\! = \begin{cases}	
		\log \rho + \mathcal{O}(1), & \alpha_2 \le 1, \\
			(\min\{2-k_t,1\}\alpha_2  + \\ \min\{[1-(2-k_t)]^+,2\})\log \rho + \mathcal{O}(1),  & 1 < \alpha_2, 
			\end{cases} \label{BE5}
\end{eqnarray}
where (a) is from Gaussian input maximizing the entropy with covariance constraints; (b) is from 
$\textbf{L}[t] \preceq \textbf{I}_2$ for $\text{tr}\{\textbf{L}[t]\} \le 1$ and the SVD of $\textbf{K}[t]$; and (c) is from \cite[Lemma 1]{21}. 

As such, the upper bound of weighted sum of achievable rates from \eqref{BE2} and \eqref{BE3} is given in \eqref{BE4}, shown on the top of next page,
\begin{figure*}
	\begin{eqnarray}
		&& \!\!\!\!\!\!\!\!\!\!\! n\left(R_1 + \frac{R_2}{2} - \epsilon_n \right)  \overset{(a)}{\le}  \sum_{t=1}^n f(N,(1,M),(\alpha_2,M-k_t)) \log \rho   + \frac{1}{2} \sum_{t=1}^n h(\overline{y}_1[t],\overline{y}_2[t]|\mathcal{V}[t]) -  \sum_{t=1}^n h(\overline{y}_1[t]|\mathcal{V}[t]) + n \mathcal{O}(1) \nonumber \\
		&& \!\!\!\!\!\!\!\!\!\!\! \qquad \quad \overset{(b)}{\le} \begin{cases}
			\sum_{t=1}^n((1+ \min\{2-k_t,1\}/2 + \min\{[1-k_t]^+,1\}\alpha_2/2   -\min\{2-k_t,1\}\alpha_2)\log \rho + \mathcal{O}(1))	, & \alpha \le 1, \\
			\sum_{t=1}^n((\min\{[1-(2-k_t)]^+,2\} +\min\{2-k_t,1\}\alpha_2/2 + \min\{[1-k_t]^+,1\}/2)\log \rho + \mathcal{O}(1)), & 1 < \alpha_2.
		\end{cases}  \nonumber \\
	&& \!\!\!\!\!\!\!\!\!\!\! \qquad \quad \overset{(c)}{\le}  \begin{cases}
		\dfrac{3-\alpha_2}{2}n\log \rho + \mathcal{O}(1), & \alpha_2 \le 1, \\	 
		\dfrac{1+\alpha_2}{2} n\log \rho + \mathcal{O}(1), & 1 < \alpha_2, \\ 
	\end{cases} \label{BE4}
	\end{eqnarray}
\hrule
\end{figure*}
where (a) is from \eqref{BE2}, \eqref{BE3} and \eqref{BE5}; (b) is from \eqref{BE6} and \eqref{BE7}; (c) is from maximizer is $k_t = 0$ by exhausting $k_t \in \{0,1,2\}$. Then, we rewrite \eqref{BE4} into GDoF expression, 
	\begin{equation}
	    d_1(\alpha_1,\alpha_2) + \frac{d_2(\alpha_1,\alpha_2)}{2} \  \le  \begin{cases}
			\dfrac{3-\alpha_2}{2}, & \alpha_2 \le 1, \\	 
			\dfrac{1+\alpha_2}{2}, & 1 < \alpha_2, \\ 
		\end{cases}  \label{BE8}
	\end{equation} 
Due to the symmetry, we have another GDoF inequality, i.e., 
	\begin{equation}
 d_2(\alpha_1,\alpha_2) + \frac{d_1(\alpha_1,\alpha_2)}{2}  \le  \begin{cases}
			\dfrac{3-\alpha_1}{2}, & \alpha_1 \le 1, \\	 
			\dfrac{1+\alpha_1}{2}, & 1 < \alpha_1, \\ 
		\end{cases}  \label{BE9}
	\end{equation}
Moreover, considering single-user GDoF bound for MIMO point-to-point channel, we have
\begin{equation}
	d_i(\alpha_1, \alpha_2) \le 1, \quad i=1,2. \label{SB1}
\end{equation}

Combing \eqref{BE8} with \eqref{BE9} and \eqref{SB1}, we derive the sum-GDoF upper bound in Theorem 1 (see \eqref{BE10}).  This ends the proof.

	\section{Achievability Proof of Theorem 1}

%According to the results of \cite{21}, it is known that the block-Markov transmission scheme is sub-optimal in weak and very weak interference regimes. Therefore, we resort to the other design of transmission scheme, given below. 
 
\subsection{Proposed Transmission Scheme for $1 < \alpha_1,\alpha_2$ and $1 < \alpha_1\,\&\,\alpha_2 \le 1 \,\&\, 2 \le \alpha_1 + 2\alpha_2$ Cases} 

	The proposed transmission scheme is with block-Markov structure, and has $B$ blocks with $s$ time slot each block. Without loss of generality, we assume $s=1$.
	
	% In the $b^{th}$ block, the transmitter $\text{Tx}_1$ sends desired signal to receiver $\text{Rx}_1$ with power $\mathcal{O}(\rho^{-A_1})$, which causes interference at power $\mathcal{O}(\rho^{\alpha_1-A_1})$ at receiver $\text{Rx}_2$. In the meantime, the transmitter $\text{Tx}_2$ sends desired signal to receiver $\text{Rx}_2$ with power $\mathcal{O}(\rho^{-A_2})$, which causes interference at power $\mathcal{O}(\rho^{\alpha_2-A_2})$ at receiver $\text{Rx}_1$. In the $(b+1)^{th}$ block, the transmitter $\text{Tx}_i$ reconstructs its interference caused at receiver $\text{Rx}_i$ and block $b$, using delayed CSIT. Then, the transmitter $\text{Tx}_i$ compresses it through quantization such that the average distortion does not exceed the noise power level, hence can be ignored in GDoF analysis. The details of the transmission scheme are elaborated below.

In the $b^{th} (1\le b < B)$ block, the transmitter $\text{Tx}_i$ encodes the message $w_{i}[b]$ desired by receiver $\text{Rx}_i$ using the vector $\textbf{u}_i(w_{i}[b]) \in \mathbb{C}^{2\times 1}$, such that $\textbf{u}_i \sim \mathcal{CN}(0,\rho^{-A_i}\textbf{I}_2)$ with $0 \le A_i \le \alpha_1$. The common message $l_{i}[b-1]$  is encoded using the vector $\textbf{x}_{ic}(l_{i}[b-1]) \in \mathbb{C}^{2\times 1}$, which is transmitted at transmitter $\text{Tx}_i$ with power $\mathcal{O}(\rho^{0})$. The transmit signal at block $b$ and transmitter $\text{Tx}_i$ can be written as follows:
\begin{equation}
	\textbf{x}_i[b] = \underbrace{\textbf{u}_i(w_{i}[b])}_{\mathcal{O}(\rho^{-A_i})} + 	\underbrace{\textbf{x}_{ic}(l_{i}[b-1])}_{\mathcal{O}(\rho^{0})},
\end{equation}
where $i=1,2$. The received signal at block $b$ and receiver $\text{Rx}_i$ is given as follows:
\begin{eqnarray}
	y_i[b] = \underbrace{\sqrt{\rho} \textbf{h}_{ii}[b]\textbf{x}_{ic}(l_i[b-1])}_{\mathcal{O}(\rho^1)} + \underbrace{\sqrt{\rho^{\alpha_j}} \textbf{h}_{ij}[b]\textbf{x}_{jc}(l_j[b-1])}_{\mathcal{O}(\rho^{\alpha_j})} \nonumber \\
	+ \underbrace{\sqrt{\rho} \textbf{h}_{ii}[b]\textbf{u}_i(w_i[b])}_{\mathcal{O}(\rho^{1-A_i})} + \underbrace{\sqrt{\rho^{\alpha_j}} \textbf{h}_{ij}[b]\textbf{u}_j(w_j[b])}_{  \eta_{i}[b] \sim \mathcal{O}(\rho^{\alpha_j-A_j})} + n_i[b],
\end{eqnarray}
where $i \ne j$, and $\eta_{i}[b]$ denotes the interference at receiver $\text{Rx}_i$, which can be reconstructed at block $b+1$. From the rate distortion theorem \cite{22}, this interference $\eta_{i}[b]$ can be quantized using a source codebook with size $\mathcal{O}(\rho^{\alpha_j - A_j})$, such that the average distortion does not exceed the noise power level and can be ignored in GDoF analysis. The quantization index of interference $\eta_{i}[b]$ is denoted by $l_j[b]$, which is transmitted as $\textbf{x}_{jc}(l_j[b])$ from transmitter $\text{Tx}_j$ in block $b+1$.  

In the $B^{th}$ block, the transmitter $\text{Tx}_i$ sends common messages only, namely
\begin{equation}
	\textbf{x}_i[B] = 	\underbrace{\textbf{x}_{ic}(l_{i}[B-1])}_{\mathcal{O}(\rho^{0})},
\end{equation}
where $i=1,2$. The received signal at block $B$ and receiver $\text{Rx}_i$ is given as follows:
\begin{eqnarray}
	&& y_i[B] =  \underbrace{\sqrt{\rho} \textbf{h}_{ii}[B]\textbf{x}_{ic}(l_i[B-1])}_{\mathcal{O}(\rho^1)} +\nonumber \\
	&& \qquad \qquad  \underbrace{\sqrt{\rho^{\alpha_j}} \textbf{h}_{ij}[B]\textbf{x}_{jc}(l_j[B-1])}_{\mathcal{O}(\rho^{\alpha_j})} + n_i[B],
\end{eqnarray}
where $i=1,2$. The decoding procedure is backward and begins with block $B$, where the common messages are firstly decoded.  After that, at the $(B-1)^{th}$ block, the interference from transmitter $\text{Tx}_j$ can be canceled and extra information about $\textbf{u}_i(w_i[B-1])$ can be provided. Generally, the equivalent channel for decoding can be written  as follows: 
\begin{eqnarray}
 \underbrace{\begin{bmatrix}
			y_i[b] - \eta_{i}[b] \\
			\eta_{j}[b]
	\end{bmatrix}}_{Y'_i} = \underbrace{\begin{bmatrix}
			\sqrt{\rho} \textbf{h}_{ii}[b] \\
			\textbf{0}
	\end{bmatrix}}_{\textbf{S}_{ic}} \underbrace{\textbf{x}_{ic}(l_i[b-1])}_{d_{\eta_j}} 
 \nonumber \\ 	+ \underbrace{\begin{bmatrix}
			\sqrt{\rho^{\alpha_j}} \textbf{h}_{ij}[b] \\
			\textbf{0}
	\end{bmatrix}}_{\textbf{S}_{jc}}
  \underbrace{\textbf{x}_{jc}(l_j[b-1])}_{d_{\eta_i}}     
 \nonumber \\ 	+ \underbrace{\begin{bmatrix}
			\sqrt{\rho} \textbf{h}_{ii}[b] \\
			\sqrt{\rho^{\alpha_i}} \textbf{h}_{ji}[b]
	\end{bmatrix}}_{\textbf{S}_i} \underbrace{\textbf{u}_i(w_i[b])}_{d_i[b]} + \begin{bmatrix}
		n_i[b] \\
		0
	\end{bmatrix}, 
	\label{LE1} 
\end{eqnarray}	
where $i,j=1,2$ and $i\ne j$. Note that \eqref{LE1} is equivalent to the three-user  multiple-access channel (MAC). Applying capacity region of three-user MAC, we have the following general condition for achievable GDoF tuple to our problem:

\textbf{Proposition 1}: For the GDoF tuple $(d_{\eta_1},d_{\eta_2},d_1[b],d_2[b])$, which denote the GDoF carried in $\textbf{x}_{2c}(l_2[b-1])$, $\textbf{x}_{1c}(l_1[b-1])$, $\textbf{u}_1(w_1[b])$, and $\textbf{u}_2(w_2[b])$, respectively, we have (19)-(29), where $f(\cdot)$ and $g(\cdot)$ are defined in \cite[Lemmas 1 \& 2]{21}.
\begin{eqnarray}
	&&  \!\!\!\!\!\!\!\!\! d_{\eta_1} \le \min\{\alpha_2, 1\}, \label{P1} \\ 
	&&  \!\!\!\!\!\!\!\!\! d_{\eta_2} \le \min\{\alpha_1, 1\}, \label{P2} \\
	&& \!\!\!\!\!\!\!\!\! d_1[b] \le f(2,(1-A_1,1),(\alpha_1 - A_1,1)), \label{P3} \\ 
	&&  \!\!\!\!\!\!\!\!\! d_2[b] \le f(2,(1-A_2,1),(\alpha_2 - A_2,1)), \label{P4}\\
	&&  \!\!\!\!\!\!\!\!\! d_{\eta_1} + d_{\eta_2} \le f(1,(1,2),(\alpha_2,2)), \label{P5} \\
	&&  \!\!\!\!\!\!\!\!\! d_{\eta_1} + d_1[b] \le   \alpha_1 - A_1 \nonumber  \\
	&& \!\!\!\!\!\!\!\!\! \qquad +g(1, (\alpha_2, 2), (1-\alpha_1, 1), (1 - A_1, 1)), \label{P6}\\
	%\alpha_1 - A_1 + f(1,(1,2),(\alpha_2,2)), \label{P6} \\
	&&  \!\!\!\!\!\!\!\!\! d_{\eta_2} + d_2[b] \le   \alpha_2 - A_2 \nonumber \\
	%\alpha_2 - A_2 + f(1,(1,2),(\alpha_1,2)), \label{P7} \\
	&& \!\!\!\!\!\!\!\!\! \qquad +g(1, (\alpha_1, 2), (1-\alpha_2, 1), (1 - A_2, 1)), \label{P7} \\
	&&  \!\!\!\!\!\!\!\!\! d_{\eta_2} + d_1[b] \le \alpha_1 - A_1 +1, \label{P8} \\
	&&  \!\!\!\!\!\!\!\!\! d_{\eta_1} + d_2[b] \le \alpha_2 - A_2  + 1, \label{P9} \\
	&&  \!\!\!\!\!\!\!\!\! d_{\eta_1} + d_{\eta_2} +  d_1[b] \nonumber \\
	&&  \!\!\!\!\!\!\!\!\! \quad \le \alpha_1 - A_1 + f(1,(1,2),(\alpha_2,2)), \label{P10} \\
	&&  \!\!\!\!\!\!\!\!\! d_{\eta_1} + d_{\eta_2} +  d_2[b] \nonumber \\
	&&  \!\!\!\!\!\!\!\!\! \quad \le \alpha_2 - A_2 + f(1,(1,2),(\alpha_1,2)), \label{P11} 	
\end{eqnarray}
\begin{IEEEproof}
 The proof is similar to that in \cite{21}. Thus, it is omitted for simplicity.
\end{IEEEproof}

%\subsubsection{Analysis of Achievable Sum-GDoF} 
 
 For the block-Markov transmission, the achievable GDoF of receiver $i$ is calculated as $d_i = \lim_{b \rightarrow {\cal{1}}} \frac{1}{B} \sum_{b=1}^B d_i[b] = d_i[B]$. Moreover, $d_{\eta_i}$ is allocated as $\alpha_j - A_j$, since the common message from transmitter $\text{Tx}_i$ need to be decoded at receiver $\text{Rx}_j$. In the following, we analyze the achievable sum-GDoF case by case, by means of Proposition 1.
 
 \subsubsection{$1 < \alpha_1,\alpha_2$ Case}
 
 According to Proposition 1, we present the achievable GDoF condition in this case as follows:
 \begin{eqnarray}
 	&& \!\!\!\!\!\! \alpha_1 - A_1 \le 1, \quad \alpha_2 - A_2 \le 1, \\ 
 	&& \!\!\!\!\!\! d_1 \le \alpha_1 + 1 -2A_1, \\
 	&& \!\!\!\!\!\! d_2 \le \alpha_2 + 1 -2A_2, \\
 	&& \!\!\!\!\!\! \alpha_1 + \alpha_2 \le A_1 + A_2 + 1, \\
 	&& \!\!\!\!\!\! d_1   \le 
 \alpha_1 -A_1 + A_2, \\
 	&& \!\!\!\!\!\! d_2    \le 
 	\alpha_2 -A_2 + A_1,\\ 
 	&&  \!\!\!\!\!\! d_1 \le 1, \\
 	&& \!\!\!\!\!\! d_2 \le 1, \\
 	&& \!\!\!\!\!\! d_1  \le A_2, \\
 	&&  \!\!\!\!\!\! d_2 \le A_1.  
 \end{eqnarray}	
 Therefore, we are able to formulate the following sum-GDoF lower bound maximization problem:
 \begin{equation}
 	\max_{A_1,A_2} \min \{A_1 + A_2, \alpha_1 + \alpha_2 + 2 -2(A_1 + A_2),2\},
 \end{equation}
 where the maximizer is $A_1^* + A_2^* = \min\{(2 + \alpha_1 + \alpha_2)/3,2\}$. This leads to the sum-GDoF lower bound $\min\{(2 + \alpha_1 + \alpha_2)/3,2\}$ achievable.

 \subsubsection{$1 < \alpha_1\,\&\, \alpha_2 < 1\,\&\,2 \le \alpha_1 + 2\alpha_2$ Case}
 
 According to Proposition 1, we present the achievable GDoF condition in this case as follows:
 \begin{eqnarray}
 	&& \!\!\!\!\!\! 0 \le A_2,  \quad \alpha_1 - A_1 \le 1, \\ 
 	&& \!\!\!\!\!\! d_1 \le \alpha_1 + 1 -2A_1, \\
 	&& \!\!\!\!\!\! d_2 \le \alpha_2 + 1 - 2A_2, \\
 	&& \!\!\!\!\!\! \alpha_1  +\alpha_2 \le 1 + A_1 + A_2, \\
 	&& \!\!\!\!\!\! d_1 \le\begin{cases}
 		\alpha_1  - A_1 + A_2, & 1 - A_1 \le \alpha_2, \\
 		\alpha_1 -\alpha_2 +A_2+ 1 -2A_1,& 1 - A_1> \alpha_2,
 	\end{cases} \\
 	&& \!\!\!\!\!\! d_2 \le\alpha_2  - A_2 + A_1, \\
 	&& \!\!\!\!\!\! d_1 \le 1, \\
 	&& \!\!\!\!\!\! d_2 \le 1, \\
 	&& \!\!\!\!\!\! d_1 \le A_2 - \alpha_2 + 1, \\
 	&& \!\!\!\!\!\! d_2 \le A_1.
 \end{eqnarray}	 
 Therefore, due to $2 \le \alpha_1 + 2\alpha_2$, we are able to formulate the following sum-GDoF lower bound maximization problem:	
 \begin{equation}
 	\max_{A_1,A_2} \min \{A_1 + A_2 - \alpha_2 + 1, \alpha_1 + \alpha_2 + 2 -2(A_1 + A_2)\},
 \end{equation}
 where the maximizer is $A_1^* + A_2^* = (\alpha_1 + 2\alpha_2 + 1)/3$. This leads to the sum-GDoF lower bound $(4 + \alpha_1 - \alpha_2) /3$ achievable.

\subsection{Proposed Transmission Scheme for $\alpha_1,\alpha_2 \le 1$ Case}

 In the $1^\text{st}$ time slot, the transmitter $\text{Tx}_1$ sends three symbols for receiver $\text{Rx}_1$ and transmitter $\text{Tx}_2$ sends one symbol for receiver $\text{Rx}_2$. 
Let us denote the symbols desired by receiver $\text{Rx}_1$ transmitted in time slot $1$ by $a_1,a_2,a_3$, and denote the symbol desired by  receiver $\text{Rx}_2$ transmitted in time slot $1$ by $b_1$. The transmit signal at transmitter $\text{Tx}_1$ is designed as 
\begin{equation}
	\textbf{x}_1[1] = \begin{bmatrix}
		a_1 \\
		a_2
	\end{bmatrix} +
\begin{bmatrix}
	a_3 \rho^{-\alpha_1/2} \\
	\phi
\end{bmatrix}.
\end{equation}
The transmit signal at transmitter $\text{Tx}_2$ is designed as
\begin{equation}
	\textbf{x}_2[1] =  
	\begin{bmatrix}
		b_1 \rho^{-\alpha_1/2} \\
		\phi
	\end{bmatrix}.
\end{equation}
As such, the received signals at each receiver  are expressed as
\begin{subequations}
	\begin{eqnarray}
&&	 y_1[1] = 	\underbrace{\sqrt{\rho}\textbf{h}_{11}[1]\begin{bmatrix}
			a_1 \\
			a_2
		\end{bmatrix}}_{\mathcal{O}(\rho^{1-(1-\alpha_1)})=\mathcal{O}(\rho^{\alpha_1})} + 
		\underbrace{\sqrt{\rho}\textbf{h}_{11}[1] \begin{bmatrix}
			a_3 \rho^{-\alpha_1/2} \\
			\phi
		\end{bmatrix}}_{\mathcal{O}(\rho^{1-\alpha_1})} \nonumber \\
&&	 \qquad \quad + \underbrace{\sqrt{\rho^{\alpha_2}}\textbf{h}_{12}[1] \begin{bmatrix}
			b_1 \rho^{-\alpha_1/2} \\
			\phi
		\end{bmatrix}}_{\mathcal{O}(\rho^{0})}, \\
&&	y_2[1] = 	\underbrace{\sqrt{\rho^{\alpha_1}}\textbf{h}_{21}[1]\begin{bmatrix}
		a_1 \\
		a_2
	\end{bmatrix}}_{\mathcal{O}(\rho^{\alpha_1})} + 
\underbrace{\sqrt{\rho^{\alpha_1}}\textbf{h}_{21}[1] \begin{bmatrix}
		a_3 \rho^{-\alpha_1/2} \\
		\phi
	\end{bmatrix}}_{\mathcal{O}(\rho^0)} \nonumber \\
&& \qquad \quad + \underbrace{\sqrt{\rho}\textbf{h}_{22}[1] \begin{bmatrix}
b_1 \rho^{-\alpha_1/2} \\
\phi
\end{bmatrix}}_{\mathcal{O}(\rho^{1-\alpha_1})}.
	\end{eqnarray}
\end{subequations}
It can be seen that a part of interference fall into noise level. Moreover, each receiver  can retrieve the profile of $\mathcal{O}(\rho^{\alpha_1})$ part and decode the information of $\mathcal{O}(\rho^{1-\alpha_1})$ part immediately.  

In the $2^\text{nd}$ time slot, the transmitter $\text{Tx}_2$ sends three symbols for receiver $\text{Rx}_2$ and transmitter $\text{Tx}_1$ sends three symbols for receiver $\text{Rx}_1$.   
Let us denote the symbols desired by receiver $\text{Rx}_2$ transmitted in time slot $2$ by $b_2,b_3,b_4$, and denote the symbol desired by  receiver $\text{Rx}_1$ transmitted in time slot $2$ by $a_4$. The transmit signal at transmitter $\text{Tx}_1$ is designed as 
\begin{equation}
	\textbf{x}_1[2] =  
	\begin{bmatrix}
		a_4 \rho^{-\alpha_2/2} \\
		\phi
	\end{bmatrix}.
\end{equation}
The transmit signal at transmitter $\text{Tx}_2$ is designed as
\begin{equation}
	\textbf{x}_2[2] = \begin{bmatrix}
		b_2 \\
		b_3
	\end{bmatrix} +
	\begin{bmatrix}
		b_4 \rho^{-\alpha_2/2} \\
		\phi
	\end{bmatrix}.
\end{equation}
As such, the received signals at each receiver  are expressed as
\begin{subequations}
	\begin{eqnarray}
	&&	y_1[2] = \underbrace{\sqrt{\rho} \textbf{h}_{11}[2] \begin{bmatrix}
			a_4 \rho^{-\alpha_2/2} \\
			\phi
		\end{bmatrix}}_{\mathcal{O}(\rho^{1-\alpha_2})} + \underbrace{\sqrt{\rho^{\alpha_2}} \textbf{h}_{12}[2] \begin{bmatrix}
		b_2 \\
		b_3
	\end{bmatrix}}_{\mathcal{O}(\rho^{\alpha_2})} \nonumber \\
&&\qquad \quad + \underbrace{\sqrt{\rho^{\alpha_2}} \textbf{h}_{12}[2] \begin{bmatrix}
	b_4 \rho^{-\alpha_2/2} \\
	\phi
\end{bmatrix}}_{\mathcal{O}(\rho^0)}, \\
&& y_2[2] = \underbrace{\sqrt{\rho^{\alpha_2}} \textbf{h}_{21}[2]\begin{bmatrix}
	a_4 \rho^{-\alpha_2/2} \\
	\phi
\end{bmatrix}}_{\mathcal{O}(\rho^{0})} + \underbrace{\sqrt{\rho} \textbf{h}_{22}[2] \begin{bmatrix}
b_2 \\
b_3
\end{bmatrix}}_{\mathcal{O}(\rho^{1-(1-\alpha_2)})= \mathcal{O}(\rho^{\alpha_2})} 
\nonumber \\
&&  \qquad \quad + \underbrace{\sqrt{\rho} \textbf{h}_{22}[2] \begin{bmatrix}
	b_4 \rho^{-\alpha_2/2} \\
	\phi
\end{bmatrix}}_{ \mathcal{O}(\rho^{1-\alpha_2})} .
	\end{eqnarray}
\end{subequations}
It can be seen that a part of interference fall into noise level. Moreover, each receiver  can retrieve the profile of $\mathcal{O}(\rho^{\alpha_2})$ part and decode the information of $\mathcal{O}(\rho^{1-\alpha_2})$ part immediately.

In the $3^\text{rd}$ time slot, each transmitter has delayed CSI of past two time slots. Thus,  the transmitter $\text{Tx}_1$ re-constructs the interfering signals and treat it as a common symbol, i.e.,
$
	c_1 \triangleq \sqrt{\rho^{\alpha_1}}\textbf{h}_{21}[1][
		a_1;
		a_2],
$
where $c_1$ is a valid codeword if $a_1,a_2$ are selected from lattice.
The transmitter $\text{Tx}_2$ re-constructs the interfering signals and treat it as a common symbol, i.e.,
$
	c_2 \triangleq \sqrt{\rho^{\alpha_2}}\textbf{h}_{12}[2][
		b_2;
		b_3
	],
$
where $c_2$ is a valid codeword if $b_2,b_3$ are selected from lattice.
The transmit signals at each transmitter are designed as
\begin{subequations}
	\begin{eqnarray}
		\textbf{x}_1[3] = \begin{bmatrix}
			c_1 \\
			\phi
		\end{bmatrix} + \begin{bmatrix}
		a_5 \rho^{-\alpha_1/2} \\
		\phi
	\end{bmatrix}, \\
		\textbf{x}_2[3] = \begin{bmatrix}
	c_2 \\
	\phi
\end{bmatrix} + \begin{bmatrix}
	b_5 \rho^{-\alpha_2/2} \\
	\phi
\end{bmatrix},
	\end{eqnarray}
\end{subequations}
where $a_5,b_5$ are symbols desired by receivers $\text{Rx}_1$ and $\text{Rx}_2$, respectively.
As such, the received signals at each receiver  are expressed as
\begin{subequations}
	\begin{eqnarray}
		&&	y_1[3] = \underbrace{\sqrt{\rho} \textbf{h}_{11}[3] \begin{bmatrix}
				c_1 \\
				\phi
		\end{bmatrix}}_{\mathcal{O}(\rho^{1-(1-\alpha_1)})=\mathcal{O}(\rho^{\alpha_1})} +  \underbrace{\sqrt{\rho} \textbf{h}_{11}[3] \begin{bmatrix}
		a_5 \rho^{-\alpha_1/2} \\
		\phi
	\end{bmatrix}}_{\mathcal{O}(\rho^{1-\alpha_1})} \nonumber \\
		&&  +  \sqrt{\rho^{\alpha_2}} \textbf{h}_{12}[3] \begin{bmatrix}
				c_2 \\
				\phi
		\end{bmatrix}  + \underbrace{\sqrt{\rho^{\alpha_2}} \textbf{h}_{12}[3] \begin{bmatrix}
		b_5 \rho^{-\alpha_2/2} \\
		\phi
	\end{bmatrix}}_{\mathcal{O}(\rho^0)}, \\
		&& y_2[3] =  \underbrace{\sqrt{\rho} \textbf{h}_{22}[3] \begin{bmatrix}
			c_2 \\
			\phi
		\end{bmatrix}}_{\mathcal{O}(\rho^{1-(1-\alpha_2)}) = \mathcal{O}(\rho^{\alpha_2})} + \underbrace{\sqrt{\rho} \textbf{h}_{22}[3] \begin{bmatrix}
				b_5 \rho^{-\alpha_2/2} \\
				\phi
		\end{bmatrix}}_{\mathcal{O}(\rho^{1-\alpha_2})}
		\nonumber \\
		&&  + \sqrt{\rho^{\alpha_1}} \textbf{h}_{21}[3] \begin{bmatrix}
				c_1 \\
				\phi
		\end{bmatrix} +  \underbrace{\sqrt{\rho^{\alpha_1}} \textbf{h}_{21}[3] \begin{bmatrix}
				a_5 \rho^{-\alpha_1/2} \\
				\phi
		\end{bmatrix}}_{\mathcal{O}(\rho^{0})},  
	\end{eqnarray}
\end{subequations}
where the impact of $\sqrt{\rho^{\alpha_2}} \textbf{h}_{12}[3][c_2;\phi]$ can be subtracted from $y_1[3]$ and the  $\sqrt{\rho^{\alpha_1}} \textbf{h}_{21}[3][c_1;\phi]$ can be subtracted from $y_2[3]$. It can be seen that a part of interference fall into noise level. Moreover,  receiver $\text{Rx}_j$ can retrieve the profile of $\mathcal{O}(\rho^{\alpha_j})$ part and decode the information of $\mathcal{O}(\rho^{1-\alpha_j})$ part immediately.

The achievable sum-GDoF is calculated as follows: Receiver $\text{Rx}_1$ acquires $1-\alpha_1$, $1-\alpha_2$, and $(1-\alpha_1) \log \rho + \mathcal{O}(1)$   immediately in time slot $1,2$ and $3$, respectively. Additionally, receiver  $\text{Rx}_1$ has $2\alpha_1 \log \rho + \mathcal{O}(1)$ via delayed CSIT. Thus, $d_1 \ge 1-\alpha_2/3$  is achievable. Likewise, $d_2 \ge 1-\alpha_1/3$  is achievable. To sum up, $d_1+d_2 \ge 2 - (\alpha_1 + \alpha_2)/3$ is achievable.
%Receiver $\text{Rx}_2$ acquires $1-\alpha_2$, $1-\alpha_1$, and $(1-\alpha_2) \log \rho + \mathcal{O}(1)$   immediately in time slot $1,2$ and $3$, respectively. Additionally, receiver  $\text{Rx}_2$ has $2\alpha_2 \log \rho + \mathcal{O}(1)$ via delayed CSIT. Thus, $d_2 \ge 1-\alpha_1/3$  is achievable. To sum up, $d_1+d_2 \ge 2 - (\alpha_1 + \alpha_2)/3$ is achievable.

\section{Conclusion}

The sum-GDoF was characterized in the asymmetry interference channel with delayed CSIT, where each transmitter has 2 antennas and each receiver has 1 antenna. In the future, it is interesting to design a better transmission scheme in $1 < \alpha_1 \,\&\, \alpha_2 < 1 \,\&\, \alpha_1 + 2\alpha_2 < 2$ Case.

	\bibliographystyle{IEEEtran}
\bibliography{GDoF}	

\begin{thebibliography}{10}
\providecommand{\url}[1]{#1}
\csname url@samestyle\endcsname
\providecommand{\newblock}{\relax}
\providecommand{\bibinfo}[2]{#2}
\providecommand{\BIBentrySTDinterwordspacing}{\spaceskip=0pt\relax}
\providecommand{\BIBentryALTinterwordstretchfactor}{4}
\providecommand{\BIBentryALTinterwordspacing}{\spaceskip=\fontdimen2\font plus
\BIBentryALTinterwordstretchfactor\fontdimen3\font minus
  \fontdimen4\font\relax}
\providecommand{\BIBforeignlanguage}[2]{{%
\expandafter\ifx\csname l@#1\endcsname\relax
\typeout{** WARNING: IEEEtran.bst: No hyphenation pattern has been}%
\typeout{** loaded for the language `#1'. Using the pattern for}%
\typeout{** the default language instead.}%
\else
\language=\csname l@#1\endcsname
\fi
#2}}
\providecommand{\BIBdecl}{\relax}
\BIBdecl

\bibitem{24}
C.~S. Vaze and M.~K. Varanasi, ``The degrees of freedom region and interference
  alignment for the {MIMO} interference channel with delayed {CSIT},''
  \emph{IEEE Trans. Inf Theory}, vol.~58, no.~7, pp. 4396--4417, 2012.

\bibitem{25}
M.~J. Abdoli, A.~Ghasemi, and A.~K. Khandani, ``On the degrees of freedom of
  three-user {MIMO} broadcast channel with delayed {CSIT},'' in \emph{Proc.
  IEEE Int. Symp. Inf. Theory (ISIT)}, St. Petersburg, Russia, 2011, pp.
  209--213.

\bibitem{26}
T.~Zhang, X.~W. Wu, Y.~F. Xu, Y.~Ge, and P.~C. Ching, ``Three-user {MIMO}
  broadcast channel with delayed {CSIT}: {A} higher achievable {DoF},'' in
  \emph{Proc. IEEE Int. Conf. Acoust., Speech, Signal Process. (ICASSP)}, 2018,
  pp. 3709--3713.

\bibitem{27}
M.~J. Abdoli, ``Feedback and cooperation in wireless networks,'' in \emph{PhD
  Dissertation, University of Waterloo}, 2012.

\bibitem{28}
T.~{Zhang} and R.~{Wang}, ``Achievable {DoF} regions of three-user {MIMO}
  broadcast channel with delayed {CSIT},'' \emph{IEEE Trans. Commun.}, vol.~69,
  no.~4, pp. 2240--2253, 2021.

\bibitem{29}
D.~T.~H. Kao and A.~S. Avestimehr, ``Linear degrees of freedom of the {MIMO}
  {X}-channel with delayed {CSIT},'' \emph{IEEE Trans. Inf Theory}, vol.~63,
  no.~1, pp. 297--319, 2017.

\bibitem{40}
R.~H. Etkin, D.~N.~C. Tse, and H.~Wang, ``Gaussian interference channel
  capacity to within one bit,'' \emph{IEEE Trans. Inf. Theory}, vol.~54,
  no.~12, pp. 5534--5562, 2008.

\bibitem{31}
J.~Chen, P.~Elia, and S.~A. Jafar, ``On the two-user {MISO} broadcast channel
  with alternating {CSIT}: A topological perspective,'' \emph{IEEE Trans. Inf
  Theory}, vol.~61, no.~8, pp. 4345--4366, 2015.

\bibitem{30}
Z.~H. Awan and A.~Sezgin, ``Secure {MISO} broadcast channel: An interplay
  between {CSIT} and network topology,'' \emph{IEEE J. Sel. Areas Inf. Theory},
  vol.~2, no.~1, pp. 121--138, 2021.

\bibitem{20}
K.~Mohanty and M.~K. Varanasi, ``The generalized degrees of freedom region of
  the {MIMO} {Z}-interference channel with delayed {CSIT},'' \emph{IEEE Trans.
  Inf. Theory}, vol.~64, no.~1, pp. 531--546, 2018.

\bibitem{21}
------, ``On the generalized degrees of freedom of the {MIMO} interference
  channel with delayed {CSIT},'' \emph{IEEE Trans. Inf. Theory}, vol.~65,
  no.~5, pp. 3261--3277, 2019.

\bibitem{32}
T.~Liu and P.~Viswanath, ``An extremal inequality motivated by multiterminal
  information-theoretic problems,'' \emph{IEEE Trans. Inf Theory}, vol.~53,
  no.~5, pp. 1839--1851, 2007.

\bibitem{22}
T.~M. Cover, \emph{Elements of information theory}.\hskip 1em plus 0.5em minus
  0.4em\relax John Wiley \& Sons, 1999.

\end{thebibliography}
\end{document}